\documentclass[12pt,preprint]{aastex}

\usepackage{epsfig}
\usepackage{lscape}
\usepackage{mathptmx} 

\newcommand{\myemail}{l.loinard@crya.unam.mx}

\newcommand{\dechms}[4]{$#1^{\rm h}#2^{\rm m}#3\mbox{$^{\rm s}\mskip-7.6mu.\,$}#4$}

\newcommand{\decdms}[4]{$#1^{\circ}#2'#3\mbox{$''\mskip-7.6mu.\,$}#4$}

\newcommand{\msec}[2]{$#1\mbox{$''\mskip-7.6mu.\,$}#2$}

\shorttitle{H66\alpha$ and He66$\alpha$ in MWC349}
\shortauthors{}

\begin{document}

\title{Expanded Very Large Array observations of the H66$\alpha$ and H\lowercase{e}66$\alpha$ \\ recombination lines toward MWC~349A}

\author{Laurent Loinard and Luis F.\ Rodr\'{\i}guez}

\affil{Centro de Radioastronom\'{\i}a y Astrof\'{\i}sica, Universidad
Nacional Aut\'onoma de M\'exico\\ Apartado Postal 3-72, 58090,
Morelia, Michoac\'an, M\'exico (\myemail)}

\begin{abstract}
We have used the greatly enhanced spectral capabilities of the Expanded Very 
Large Array to observe both the 22.3 GHz continuum emission and the 
H66$\alpha$ recombination line toward the well-studied Galactic emission-line
star MWC 349A. The continuum flux density is found to be 411 $\pm$ 41 mJy 
in good agreement with previous determinations. The H66$\alpha$ line peak
intensity is about 25 mJy, and the average line-to-continuum flux ratio is about 
5\%, as expected for local thermodynamic equilibrium conditions. This shows that 
the H66$\alpha$ recombination line is not strongly masing as had previously been 
suggested, although a moderate maser contribution could be present. The 
He66$\alpha$ recombination line is also detected in our observations; the 
relative strengths of the two recombination lines yield an ionized helium to 
ionized hydrogen abundance ratio $y^+$ = 0.12 $\pm$ 0.02. The ionized 
helium appears to share the kinematics of the thermally excited
ionized hydrogen gas, so the two species are likely to be well mixed.
The electron temperature of the ionized gas in MWC 349A deduced 
from our observations is 6,300 $\pm$ 600 K.
\end{abstract}


\section{Introduction}

MWC 349A is a peculiar B[e] star which happens to be the brightest continuum 
source of its kind at cm wavelengths (e.g.\ Braes et al.\ 1972; Tafoya et al.\ 2004). 
It is located in the direction of the Cyg OB2 association, at a distance which has 
been estimated between 1.2 kpc (Cohen et al.\ 1985) and 1.7 kpc (Hofmann et al.\ 
2002). MWC349A may have a B0 III companion (named MWC 349B), located 
$2\rlap.{''}4$ to its west  (Merrill et al.\ 1932; Cohen et al.\ 1985). The cm radio 
continuum emission from MWC 349A can be successfully described in terms of 
an isothermal wind photoevaporating from the surface of a neutral Keplerian disk 
surrounding a young massive star (Lugo et al. 2004). 

MWC 349A also exhibits hydrogen recombination line emission detectable from 
the radio (Altenhoff et al.\ 1981) to the optical (Hartmann et al.\ 1980). The 
intensity of the recombination lines at cm wavelengths was found to be consistent
with nearly thermal, LTE conditions (e.g.\ Altenhoff et al.\ 1981; Escalante et al.\ 
1989; Rodr\'\i guez \& Bastian 1994). However, in a remarkable discovery, 
Mart\'\i n-Pintado et al.\ (1989a) found that the mm recombination lines towards
MWC 349A were emitting in a maser mode, and were therefore tens of times 
stronger than expected for thermal conditions. This discovery was rapidly confirmed,
and it is now known that all the hydrogen recombination lines at mm, sub-mm, far-IR 
and mid-IR wavelengths (from H39$\alpha$ at about 100 GHz to H7$\alpha$ at
about 16,000 GHz) are masing (Gordon et al.\ 2001; Thum et al.\ 1994a; Strelnitski et 
al.\ 1996a, 1996b; Thum et al.\ 1998). These maser lines are also known to be time 
variable (Mart\'\i n-Pintado et al.\ 1989b; Gordon et al.\ 2001). To this date, MWC 
349A remains the only source known to show such high-gain natural hydrogen 
maser emission. The line amplification has been investigated theoretically by Thum et 
al.\ (1994b) and Ponomarev et al.\ (1994), and is believed to arise in the inner region 
of a nearly edge-on disk where the eastern region is redshifted with respect to the 
western one (Planesas et al.\ 1992; Weintroub et al.\ 2008).

In an unexpected turn of events, Mart\'\i n-Pintado et al.\ (1993) reported that the
H66$\alpha$ recombination line at 22 GHz was about 5 times stronger than expected 
from thermal emission, and argued that this line must, therefore, also be masing. This 
was surprising and implied that the maser effect could still be present at wavelengths 
as large as 1.3 cm. The observations reported by Mart\'\i n-Pintado et al.\ (1993), 
however, were taken with the Very Large Array (VLA) at the time when it only had 
limited spectral capabilities. In particular, the observations were made with a total 
velocity coverage of only about 160 km s$^{-1}$, barely enough to include the entire
width of the H66$\alpha$ line. This restriction could have led to an improper continuum 
subtraction and to an overestimate of the line intensity. In this {\it Letter}, we take advantage 
of the greatly enhanced spectral capabilities of the Expanded Very Large Array (EVLA) to 
reobserve this important source and obtain accurate H66$\alpha$ line parameters.

\section{Observations}

The data were collected with the Expanded Very Large Array in its most compact (D) 
configuration on June 16, 2010 while 22 antennas were available in the array. Five
of these antennas had to be flagged because they consistently delivered low visibility
amplitudes or because they exhibited significant phase jumps. As a consequence,
the final data products were constructed from observations collected on 17 antennas
only. The correlator was set up to deliver 256 spectral channels, each 500 kHz wide, 
centered on the rest frequency of the H66$\alpha$ line at 22.364 GHz. This
corresponds to a total velocity coverage in excess of 1,700 km s$^{-1}$ (about 10 times 
the velocity coverage of Mart\'\i n-Pintado et al.\ 1993), and a spectral resolution of 
about 6.7 km s$^{-1}$. The data were calibrated with the Common Astronomy Software 
Applications (CASA) package following standard procedures. The bandpass
was measured using observations of J1743--0350 and J0319$+$4130 (3C84) obtained
at the beginning and the end of the observations. The flux scale was set using an 
observation of J0137$+$3309 (3C48). Finally, the amplitude and phase gains were 
monitored using observations of J2015$+$3710 (with a bootstrapped flux density of 2.3 Jy) 
obtained roughly every 10 minutes throughout the 3 hours of observations. 

Both the continuum and the H66$\alpha$ line were clearly detected in the observations, 
so two data sets were created from the calibrated visibilities. A continuum data set
was generated by averaging 137 line-free spectral channels (corresponding to 
an equivalent bandwidth of about 70 MHz). Simultaneously, a  continuum-free 
spectral line data set was created by subtracting a constant continuum (measured over
the same 137 line-free channels) from the initial data. Both data sets were then
imaged using robust weighting and 0.5 arcsecond pixels. The resulting angular
resolution was \msec{3}{35} $\times$ \msec{2}{07}; P.A.\ --60.7$^\circ$. To improve
the overall calibration, the continuum data set was self-calibrated --in phase 
only-- and the resulting gain corrections were applied both to the continuum
and the continuum-free spectral line data sets which were then re-imaged. 

\section{Results}

The continuum image shows the presence of a strong point-like
source at the expected position of MWC 349A ($\alpha_{J2000.0}$ = 
\dechms{20}{32}{45}{52}; $\delta_{J2000.0}$ = \decdms{40}{39}{36}{62}). A 2D 
Gaussian fit to this source yields an integrated flux of 411 mJy with a formal error of 
a fraction of a mJy. This error, however, does not account for the systematic uncertainty 
on the absolute flux calibration. We estimate this systematic uncertainty to be about 
10\%, and conclude that the 22.3 GHz flux of MWC 349A is $F_\nu$ = 411 $\pm$ 41 
mJy. This value is in good agreement with the flux obtained in previous 1.3 cm observations 
of MWC 349A. For instance, Tafoya et al.\ (2004) reported a 22.4 GHz flux density of 
446.2 $\pm$ 44.8 mJy. We note, also, that changes in the morphology of MWC 349A at 
cm wavelengths are known to have occurred (Rodr\'{\i}guez et al.\ 2007), so moderate
variations in its flux density might also be present.

In the continuum-free spectral data set, emission is clearly detected in more than 40 
contiguous spectral channels around the frequency of the H66$\alpha$ line. This 
emission remains unresolved at the angular resolution of our observations, so a 
single-pixel spectrum at the peak position of the source (Figure 1) provides a complete 
description of the line emission. This spectrum shows a very prominent double-peaked 
feature surrounded by a 1,200 km s$^{-1}$ emission-free spectral baseline. 
This baseline is remarkably linear but not flat. Instead, it shows a slope which most certainly 
results from the combined effect of the true positive spectral index of MWC 349A (e.g.\
Tafoya et al.\ 2004) and the uncorrected negative spectral indices of the flux and
bandpass calibrators.\footnote{In the future, CASA will be able to take into account the
spectral dependences of the calibrators, and it will become possible to measure directly
the spectral index of the target from observations such as those presented here.}
A spectrum with a linear baseline removed  is shown in Figure 1b. The VLA observations 
reported by Mart\'\i n-Pintado et al.\  (1993) only covered the spectral bandwidth shown 
by the thick horizontal bracket on Figure 1b. It is clear that this spectral window barely contains 
the most prominent spectral line detected here, and does not include at all the weaker 
feature seen at more negative velocities. Since observations with such a limited velocity 
coverage would contain no channels free of line emission, it would clearly be nearly 
impossible to properly subtract a continuum from them.

\begin{figure}[!ht]
\begin{center}
\includegraphics[width=0.65\linewidth,angle=0]{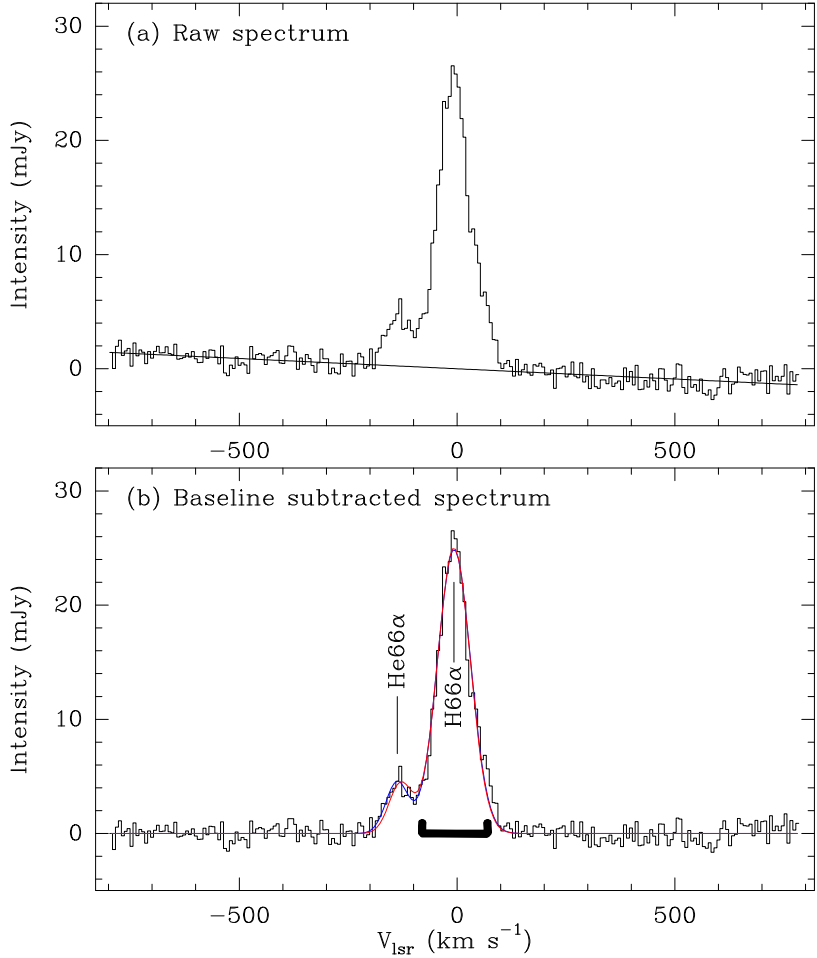}
\end{center}
\caption{(a) Raw spectrum of MWC 349A centered on the rest frequency of the H66$\alpha$ 
line. The best fit to the slope in the spectral baseline with a straight line is shown as a solid
black line. (b) Baseline-subtracted spectrum.
The velocity coverage of the VLA data presented by  Mart\'\i n-Pintado et al.\ (1993) is shown
as a horizontal bracket below the spectral lines. The blue solid line shows a fit with two unconstrained 
Gaussian components. The red solid line shows a similar fit but with the velocity separation
between the two components constrained to be 122.2 km s$^{-1}$, the expected velocity 
separation between the H66$\alpha$ and the He66$\alpha$ lines. }
\end{figure}

A simultaneous fit to the spectrum with two Gaussian components (blue solid line in Figure 1b) provides 
the parameters listed in Table 1. The separation in velocity between the two spectral features
is 130.4 $\pm$ 4.4 km s$^{-1}$, very similar to the separation (122.2 km s$^{-1}$) expected 
between the hydrogen and helium 66$\alpha$ recombination lines. Indeed, a fit where the 
separation between the two spectral components is constrained to be 122.2 km s$^{-1}$
(red solid line in Figure 1b) is very nearly as good as a fit where the separation can vary freely 
(the parameters corresponding to that constrained fit are also given in Table 1). We conclude 
that the weaker spectral feature seen at more negative velocities in our spectrum is the 
He66$\alpha$ line. We note that Thum et al.\ (1992) also identified a blue-shifted feature in 
their H41$\alpha$ spectrum of MWC 349A, and favored an interpretation where that additional 
component is the He41$\alpha$ line. The present results clearly lend strong support to that 
interpretation. Interestingly, however, the He41$\alpha$ line detected by Thum et al.\ 
(1992) appears to be at a somewhat more negative radial velocity than its hydrogen counterpart
($V_{lsr}$ = --31 km s$^{-1}$ vs.\ $V_{lsr}$ = $+$1 km s$^{-1}$). If our unconstrained fit is taken at
face value, the He66$\alpha$ line is also at a more negative radial velocity than the H66$\alpha$ 
line ($V_{lsr}$ = --15.2 $\pm$ 4.3 km s$^{-1}$ vs.\ $V_{lsr}$ = --7.0 $\pm$ 0.7 km s$^{-1}$), but 
the difference is significantly less than for the 41$\alpha$ lines. We will come back to that point in 
Section 4.3. We note, finally, that the H66$\alpha$ line also shows a hint of excess emission on 
the redshifted side that may be related to the winds predicted in the models of Avalos \& Lizano 
(in preparation).

\begin{table}[!t]
\caption{Spectral line parameters}
\begin{center}
\begin{tabular}{lrrr}
\hline
\hline
 & \multicolumn{1}{c}{$S_\nu$ (mJy)} & \multicolumn{1}{c}{$V_{lsr}$ (km s$^{-1}$)} & \multicolumn{1}{c}{$\Delta V$ (km s$^{-1}$)\tablenotemark{a}}\\%
\hline
\multicolumn{1}{l}{Unconstrained fit:}\\%
Feature 1 & 24.9 $\pm$ 0.3 & --7.0 $\pm$ 0.7 & 90.0 $\pm$ 1.7 \\%
Feature 2 & 4.5 $\pm$ 0.5 & --137.4 $\pm$ 4.3 & 61.3 $\pm$ 9.9 \\%
\multicolumn{1}{l}{Constrained fit:}\\%
H66$\alpha$ & 25.0 $\pm$ 0.3 & --7.1 $\pm$ 0.7 & 88.7 $\pm$ 1.6 \\%
He66$\alpha$ & 4.4 $\pm$ 0.4 & \multicolumn{1}{c}{$''$} & 62.8 $\pm$ 8.2 \\%
\hline
\hline
\tablenotetext{a}{$\Delta V$ is the full width at half maximum.}
\end{tabular}
\end{center}
\end{table}

\section{Discussion}

\subsection{Excitation of the H66$\alpha$ line}

The H66$\alpha$ line to continuum ratio reported by Mart\'\i n-Pintado et al.\ (1993) 
is typically about 20\%. In contrast, the line peak temperature found here (25 mJy) is 
only about 6\% of the underlying continuum. A somewhat more robust line to 
continuum ratio of about 5\% is obtained by averaging over the central 90 km 
s$^{-1}$ of the line. In LTE conditions and adopting a full width at half power of 
90 km s$^{-1}$ and an electron temperature of 8,000 K, the H66$\alpha$ line is 
predicted to represent about 5\% of the underlying continuum (Altenhoff et al.\ 
1981; Rodr\'\i guez 1982; Mart\'\i n-Pintado 2002) in excellent agreement with the 
observed values. We conclude that the H66$\alpha$ line in MWC 349A is largely 
thermally excited and is not strongly masing as had been suggested by 
Mart\'\i n-Pintado et al.\  (1993).

One could argue, however, that our results and those of Mart\'\i n-Pintado et al.\ (1993)
are not necessarily incompatible because the two sets of observations might trace 
different regions of MWC 349A. In particular, the present observations were obtained 
in the most compact (D) configuration of the EVLA and are, therefore, sensitive to even 
fairly extended emission. If it existed, such extended emission would have been filtered out in the 
observations reported by Mart\'\i n-Pintado et al.\  (1993), because they were collected in 
the most extended (A) configuration of the VLA. Could our observations be dominated 
by extended, thermally excited gas, and those of Mart\'\i n-Pintado et al.\ (1993) by denser, 
more compact masing regions? 

This appears not to be the case for the following reasons. At a wavelength of 1.3 cm, 
the (E)VLA in the A configuration is sensitive to any structure more compact than about 
2.5 arcseconds\footnote{http://evlaguides.nrao.edu/index.php?title=Category:Status}. 
Our observations, on the other hand, show that the line emission 
detected here originates in a region more compact than about 1.5 arcseconds. Thus, 
there are no spatial scales contributing to the emission reported here and to which the 
observations by Mart\'\i n-Pintado et al.\  (1993) were not sensitive. {\em The two sets of
observations trace essentially the same gas.} Moreover, at 1.3 cm, the size of MWC 
349A is about \msec{0}{2} $\times$  \msec{0}{2} (e.g.\ Tafoya et al.\  2004;  Rodr\'{\i}guez 
et al.\ 2007; Mart\'\i n-Pintado et al.\ 1993). Since the angular resolution of the (E)VLA 
at 1.3 cm in the A configuration is about \msec{0}{1}, MWC 349A only contains 4 independent 
resolution elements at that wavelength. Thus, the four spectra reported by Mart\'\i n-Pintado 
et al.\  (1993) in their Figure 2 trace most of the extent of the source, and the continuum emission 
integrated over these four positions will be about 300 mJy (the total continuum emission 
estimated by Mart\'\i n-Pintado et al.\ 1993). According to the spectra shown in Figure 2 
of Mart\'\i n-Pintado et al.\ (1993), the mean intensity of the H66$\alpha$ line over the 
velocity range from $-$60 to $+$50 km s$^{-1}$ is about 20\% of the underlying 
continuum. The line intensity integrated over these 4 positions must therefore be
about 60 mJy. As a consequence, the integrated line flux measured over any larger area should 
be at least 60 mJy in that velocity range.\footnote{Recall that flux densities measured 
in Jy do not get diluted: if a 1 Jy source is present at the center of a given field, the total 
flux density taken over any area containing the source will be 1 Jy, independently of 
how large the chosen area might be.} Yet, in our spectrum (corresponding to a 
\msec{3}{35} $\times$ \msec{2}{07} beam), the flux density is systematically smaller 
than 25 mJy; averaged over the $-$60 to $+$50 km s$^{-1}$ velocity range, it is 18 
mJy. We conclude that Mart\'\i n-Pintado et al.\  (1993) have overestimated the line 
intensity by a factor 3 to 4. 

We can only speculate about the exact reasons which led Mart\'\i n-Pintado et al.\  
(1993) to over-estimate the H66$\alpha$ line intensity in MWC 349A. In principle, 
the lack of line-free channels in their observations should have led them to over-estimate 
the continuum flux (since the channels on the edge of the bandpass used to calculate 
the continuum still contained line emission) and to under-estimate the line to continuum 
ratio (rather than the other way around). Instead, a more subtle effect, also related 
to the narrow bandwidth of their observations might have been the culprit. The velocity 
channels on the edge of the VLA bandpass are well known to be very noisy. A
continuum measurement based on these channels may have seriously underestimated
the continuum flux density, and  adversely affected the line to continuum ratio 
measurement.

\subsection{Helium abundance and electron temperature}

Our results show that the excitation of the H66$\alpha$ line
very nearly corresponds to LTE conditions. In this situation, the line behaves 
as if it were approximately optically thin (see Rodr\'{\i}guez 1982). Given its 
similar excitation and lower abundance, the He66$\alpha$ line will also 
behave as if it were approximately optically thin, so the integrated line intensity 
ratio provides a direct measure of the ionized helium over ionized hydrogen 
abundance ratio $y^+$. From the unconstrained fit  (Section 3), we obtain 
$y^+ = 0.12 \pm 0.02$, in good agreement with the value (0.10 $\pm$
0.01) found by Thum et al.\ (1992) for the 41$\alpha$ lines.

As a consequence of its relative abundance, the opacity of the helium lines is 
expected to be about 10 times less than that of the hydrogen lines, so the helium 
lines are not anticipated to be masing (even at mm wavelengths). This presumably 
explains why the helium lines associated with strongly masing hydrogen lines at
mm, sub-mm, far-IR and mid-IR wavelengths have not been detected in existing 
observations. Since the hydrogen lines are tens of times stronger than thermal, 
the helium to hydrogen line intensity ratio will typically be less than 1\% at mm 
wavelengths. The noise level of the mm observations was almost certainly adapted 
to detect the hydrogen lines, and was insufficient to reach the much weaker helium 
lines.

The high quality of the data presented here allows us to estimate the electron 
temperature of the ionized gas in MWC 349A. Following Rodr\'\i guez et al.\ 
(2009), and assuming LTE conditions, the electron temperature, $T_e^*$, of 
an ionized, partially optically-thick outflow is given by: 

\begin{equation}\Biggl[{{T_e^*} \over {K}}\Biggr] = \Biggl[{4700 \over (1 + y^+)} 
{{S_C} \over {S_L}} \biggl({{\nu_L} 
\over {GHz}} \biggr)^{1.1}
 \biggl({{\Delta v} \over 
{km~s^{-1}}}\biggr)^{-1} \Biggr]^{0.87}, \end{equation} 

\noindent where $\nu_L$ is the line frequency, $S_C$ is the continuum flux 
density, $S_L$ is the hydrogen peak line flux density, and $\Delta v$ is the hydrogen 
FWHM line width. Using the helium abundance estimated above and the line
and continuum parameters measured in Section 3, we obtain $T_e^* = 6,300 \pm 600$ K. 
This value is consistent with those determined from other radio observations
(White \& Becker 1985; Cohen et al.\ 1985; Escalante et al.\ 1989).

\subsection{Relative kinematics of the hydrogen and helium lines}

The kinematics of the ionized gas associated with MWC 349A presents several 
problems that require additional research. For instance, the direction of rotation 
determined from the cm thermal lines (eastern side blueshifted and western side 
redshifted; Rodr\'\i guez \& Bastian 1994) is opposite to that found in the mm maser 
lines (Planesas et al.\ 1992; Weintroub et al.\ 2008). Also, while the thermal radio 
recombination lines at cm wavelengths peak at negative velocities (typically
$V_{lsr}$ $\sim$ --20 km s$^{-1}$; e.g.\ Rodr\'\i guez \& Bastian 1994), the centroid 
of maser mm radio recombination lines is at positive velocities ($V_{lsr}$ $\sim$ 
$+$10 km s$^{-1}$; e.g.\ Weintroub et al.\ 2008). 

In our unconstrained fit, the LSR velocity of the He66$\alpha$ line appears to 
be --15.2 $\pm$ 4.3 km s$^{-1}$ whereas the LSR velocity of the He41$\alpha$ line 
reported by Thum et al.\ (1992) is --31 $\pm$ 8 km s$^{-1}$. Thus, on average, 
the Helium lines appear to peak at $V_{lsr}$ $\sim$ --20 km s$^{-1}$, similar to 
the velocity of the thermal cm lines of hydrogen. This strongly suggests that the helium 
lines originate in the same gas as the thermal cm hydrogen lines. 

The H66$\alpha$ line detected here peaks at a velocity
(--7.0 $\pm$ 0.7 km s$^{-1}$ -- Section 2), intermediate between the $+$10 km s$^{-1}$
masing mm lines, and the --20 km s$^{-1}$ thermal cm lines. 
A plausible (albeit somewhat {\it ad hoc}) explanation is that the H66$\alpha$
line is largely dominated by a thermal component at --20 km s$^{-1}$ but also contains a 
moderate masing component around $+$10 km s$^{-1}$. The reversed situation might 
occur with the H41$\alpha$ line at 92 GHz reported by Thum et al.\ (1992) which peaks 
at $+$1.0 $\pm$ 0.9 km s$^{-1}$. In that second case, the emission might be dominated 
by the masing line at $+$10 km s$^{-1}$ but with a moderate thermal contribution. It would 
be interesting to analyze further the relative kinematics of the hydrogen and helium lines, 
for instance by obtaining high resolution EVLA observations of the 53$\alpha$ and 
66$\alpha$ lines at 0.7 and 1.3 cm, respectively. 

\section{Conclusions and perspectives}

We have presented new EVLA observations of the 22.3 GHz continuum and the
H66$\alpha$ line emissions toward the well-studied Galactic emission-line star MWC 349A. 
The continuum flux density (411 $\pm$ 41 mJy) is in good agreement with previous 
determinations. The intensity of the H66$\alpha$ line (25 mJy) and the line-to-continuum 
flux density ratio (5\%), on the other hand, are those expected for local thermodynamic 
equilibrium. This shows that the H66$\alpha$ line excitation is largely thermal, and that 
the line is not strongly masing as had been proposed by Mart\'\i n-Pintado et al.\  (1993). 
The intermediate radial velocity of the H66$\alpha$ line, however, might suggest the 
existence of a moderate maser component. The He66$\alpha$ recombination line is also 
detected in our observations, and shows that the ionized helium to ionized hydrogen 
abundance ratio in MWC 349A is $y^+$ = 0.12 $\pm$ 0.02. Because of their shared 
kinematics, it is likely that the ionized helium gas and the thermally excited ionized 
hydrogen gas are well mixed. The electron temperature of that gas is 6,300 $\pm$ 600 K. 

High-resolution EVLA observations of this and other recombination lines (particularly
the H53$\alpha$ at 0.7 cm) might help resolve some of the pending issues concerning 
the kinematics of the gas in MWC 349A. On a more general note, we would like to stress 
that the present observations demonstrate the enormous potential of the EVLA to observe 
wide spectral lines. When previous VLA observations had to rely on {\it ad hoc} and 
often very uncertain procedures to remove a spectral baseline, the present observations 
provide a very clean identification of the line and extremely reliable determinations of
the line parameters.

\acknowledgments We thank Clemens Thum and an anonymous referee for 
thoughtful comments on our manuscript, and Vivek Dhawan, Lorant Sjouwerman, 
Rick Perley, George Moellenbrock, and Michael Rupen for their help and advices 
at various stage of the data calibration. We acknowledge the support of DGAPA, 
UNAM, and of CONACyT (M\'exico). LL is indebted to the Guggenheim Memorial 
Foundation for financial support.

\end{document}